\documentclass[lettersize,journal,twoside]{IEEEtran}
\usepackage{amsmath,amsfonts}
\usepackage{algorithmic}
\usepackage{algorithm}
\usepackage{array}
\usepackage[caption=false,font=normalsize,labelfont=sf,textfont=sf]{subfig}
\usepackage{textcomp}
\usepackage{stfloats}
\usepackage{url}
\usepackage{verbatim}
\usepackage{graphicx}
\usepackage{cite}
\usepackage{multirow}
\usepackage{bm}
\usepackage{upgreek}
\usepackage{amssymb}
\begin{document}

	\title{A new Speech Feature Fusion method with cross gate parallel CNN for Speaker Recognition}

	\author{Jiacheng Zhang, Wenyi Yan, Ye Zhang
		\thanks{This work was supported by the National Nature Science Foundation of China under Grant 61866024. {\itshape(Corresponding author: Ye Zhang.)}
			
			J. Zhang, W. Yan and Y. Zhang are with the Department of Electronic and Information Engineering Nanchang University, Nanchang 330031, China (e-mail: zhangye@ncu.edu.cn)}

	}

	
	\maketitle

	\begin{abstract}
In this paper, a new  speech feature fusion method is proposed for speaker recognition on the basis of the cross gate parallel convolutional neural network (CG-PCNN).
The Mel filter bank features (MFBFs) of different frequency resolutions can be extracted from each speech frame of a speaker's speech by several Mel filter banks, where the numbers of the triangular filters in the Mel filter banks are different.
Due to the frequency resolutions of these MFBFs are different, there are some complementaries for these MFBFs. The CG-PCNN is utilized to extract the deep features from these MFBFs, which applies a cross gate mechanism to capture the complementaries for improving the performance of the speaker recognition system.
Then, the fusion feature can be obtained by concatenating these deep features for speaker recognition.
The experimental results show that the speaker recognition system with the proposed speech feature fusion method is effective, and marginally outperforms the existing state-of-the-art systems.	
	\end{abstract}
	
	
	\section{Introduction}
	\IEEEPARstart{S}peaker recognition aims to recognize the identity of a speaker by the speaker's speech and has been widely used in many areas, such as forensics \cite{J1, J52}, general business interactions \cite{J2, J47}, and forensic voice verification \cite{J60}.
	Generally, the speaker recognition can be categorized into speaker identification and speaker verification \cite{J45}. The speaker identification is to recognize the identity of a speaker from a given set of speakers by the speaker's speech. The speaker verification aims to accept or reject the claimed identity of a speaker by the speaker's speech. This paper mainly focuses on the speaker identification.
	
	Generally, the acoustic features can be extracted from a speaker's speech by some feature extraction methods for speaker recognition, such as Mel frequency cepstral coefficients (MFCC) \cite{J6, J7}, gammatone frequency cepstral coefficients (GFCC) \cite{J33} and MFBF \cite{J34, J25, J55, J54}, etc. In \cite{J58}, the MFCCs of a speaker's speeches are employed to construct the speaker's model by the gaussian mixture model technique for speaker recognition. In \cite{J21},
	the i-vector feature is extracted by the total variability algorithm with the mean and variance normalized GFCC of a speaker's speech, then the i-vector feature is fed into the speaker's probabilistic linear discriminant analysis (PLDA) model for speaker identification. Using the MFBF of a speaker's speech \cite{J28}, a multi-layer perceptron is trained  to extract the d-vector feature of the speaker's speech for  speaker verification. In \cite{J23}, the x-vector feature is extracted from the MFBF of a speaker's speech by a deep neural network (DNN) for  speaker recognition, where the DNN is a time delay neural network (TDNN) with a statistics pooling layer. In \cite{J16}, the spectrogram feature of a speaker's speech is utilized as the input of a convolutional neural network (CNN) to extract the deep feature of the speaker's speech, then the deep feature is fed into the softmax layer for speaker identification.
	
	Generally, different types of the acoustic features of a speaker's speech can be incorporated as the fusion feature for improving the performances of the speaker recognition systems \cite{J43, J27, J32}.
	In \cite{J43}, the fusion feature of a speaker's speech is obtained by the concatenation of the MFCC and the chroma energy normalized statistics (CENS) feature of the speaker's speech, then the fusion feature is employed as the input of a CNN for speaker identification.
	In \cite{J27}, the fusion feature of a speaker's speech  is obtained with the MFBF and the Cochleagram feature of the speaker's speech by the harmonic mean calculation method, and employed as the input of a CNN with a asymmetric bi-directional long short-time memory network to obtain the deep feature of the speaker's speech for speaker identification.
	In \cite{J32}, the MFCC and the linear predictive coding (LPC) of a speaker’s speech are fed into the 1D-Triplet CNN to extract the fusion feature for speaker verification.
	\begin{figure*}[!t]
		\centering
		\includegraphics[width=5.2in]{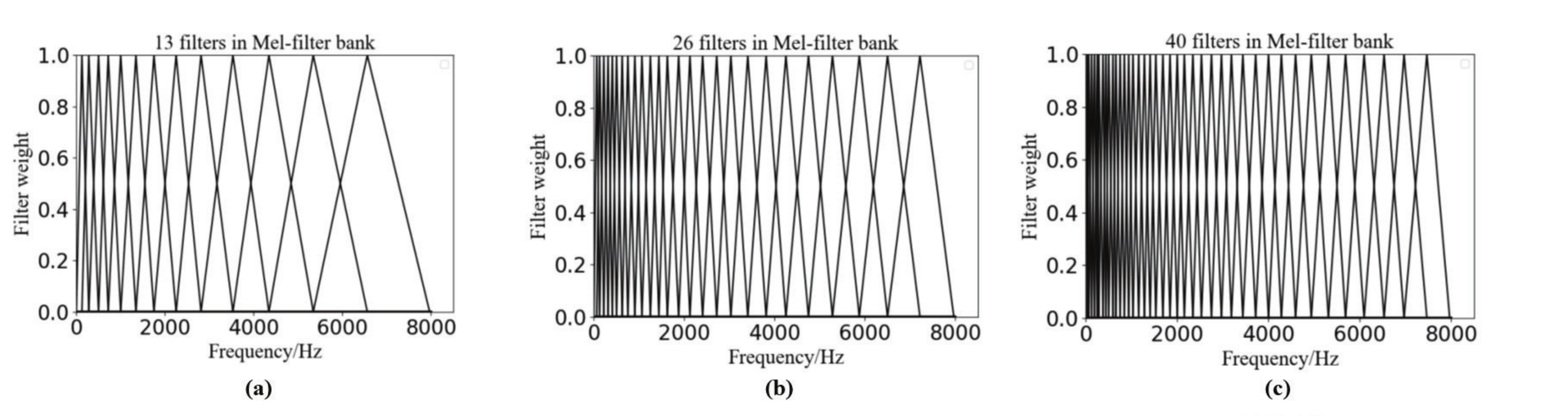}
		\caption{The frequency domain response of the Mel filter banks. (a), (b) and (c) are the frequency domain responses of the Mel filter banks, respectively, where the number of the triangular filters in the Mel filter banks are 13, 26 and 40.}
		\label{f3}
	\end{figure*}
	\begin{figure*}[!t]
	\centering
	\includegraphics[width=5.2in]{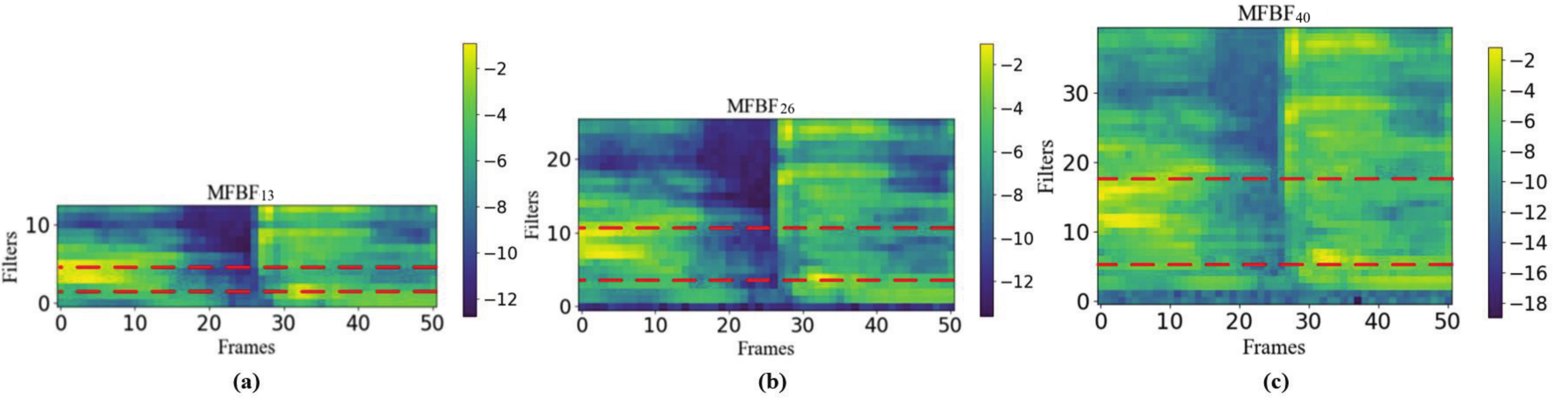}
	\caption{The MFBFs of a speaker's speech. (a), (b), (c) are the MFBF$_{13}$, MFBF$_{26}$, MFBF$_{40}$ of the speaker's speech, respectively. The area between the two red dashed lines represents approximately the 320 Hz $ \sim $ 1500 Hz frequency range.}
	\label{f4}
	\end{figure*}
	Based on the DNN, the gate mechanism  also can be applied in speaker recognition \cite{J41, J38}. In \cite{J41}, the conditional generative adversarial network (CGAN) is used for data augmentation, and the CNN with the gate mechanism is utilized in the generation network of the CGAN to generate the fake acoustic features by imitating the MFBFs of the speakers' speeches. Then the fake acoustic features are used to augment the training set and train the speaker recognition model for speaker identification. In \cite{J38}, the gate mechanism is utilized in each block of the residual network to extract a deep feature from the spectrogram feature of a speaker's speech, then the deep feature is fed into the softmax layers for speaker recognition and other audio classification tasks.
	
	In this paper, we proposed a speech feature fusion method based on the CG-PCNN for speaker recognition.
	For each speech frame of a speaker's speech, two MFBFs of different frequency resolutions are extracted by two Mel filter banks, where the numbers of the triangle filters in the Mel filter banks are different. As the number of the triangle filters in the Mel filter bank is increased, the bandwidth of each triangle filter is decreased, and more details in the frequency domain can be obtained by the Mel filter bank. Thus, some complementaries about frequency information exists within these MFBFs.
	Then, the CG-PCNN is used to extract the deep features from these MFBFs, which utilizes a cross gate mechanism to capture the complementaries within these MFBFs for improving the performance of the speaker recognition system.
	Finally, the deep features extracted by the CG-PCNN are concatenated as the fusion feature for speaker recognition.
	
	The contributions of this paper can be summarized as follows:
	
	$\bullet $ We proposed a new speech feature fusion method to fuse the MFBFs of different frequency resolutions of a speaker's speech for speaker recognition.
	
	$\bullet $ The cross gate parallel convolutional neural network (CG-PCNN) is proposed for speech feature fusion and applied in a speaker recognition system. In the CG-PCNN, the cross gate mechanism is employed to obtain the complementaries within the MFBFs of different frequency resolutions for improving the performance of the speaker recognition system.
	
	The outline of this paper is as follows. Section \uppercase\expandafter{\romannumeral2} illustrates the details of the MFBF. The proposed  feature fusion method is introduced in Section \uppercase\expandafter{\romannumeral3}. Experiments on the AISHELL-1 dataset \cite{J50}, Librispeech dataset \cite{J53} and Voxceleb2 dataset \cite{J26} are presented in Section \uppercase\expandafter{\romannumeral4}. Conclusions are given in the last section.
	
	\section{Mel Filter Bank Feature}
	The MFBF can be extracted from a speech frame of a speaker's speech by a Mel filter bank. The Mel filter bank consists of $M$ triangular filters and the frequency domain response of the Mel filter bank is described as follows:
	\begin{equation}
		\centering
		\begin{array}{l}
			{H_m}(k) = \left\{ {\begin{array}{*{20}{l}}
					{0,}&{k < f(m - 1)}\\
					{\frac{{k - f(m - 1)}}{{f(m) - f(m - 1)}},}&{f(m - 1) < k < f(m)}\\
					{\frac{{f(m + 1) - k}}{{f(m + 1) - f(m)}},}&{f(m) < k < f(m + 1)}\\
					{0,}&{k > f(m + 1)}
			\end{array}} \right.,\\
			\quad \quad \quad \quad \quad \quad\quad \quad \quad \quad \quad \quad \quad \quad \quad m = 1,2, \ldots ,M
		\end{array}
	\end{equation}
	where $M$ is the number of the triangular filters in the Mel filter bank, ${k}$ is the frequency value. $f(m - 1)$ and $f(m + 1)$ are the low and high cutoff frequencies of ${m^{th}}$ triangular filter, respectively.
	\begin{figure*}[!t]
		\centering
		\includegraphics[width=5.5in]{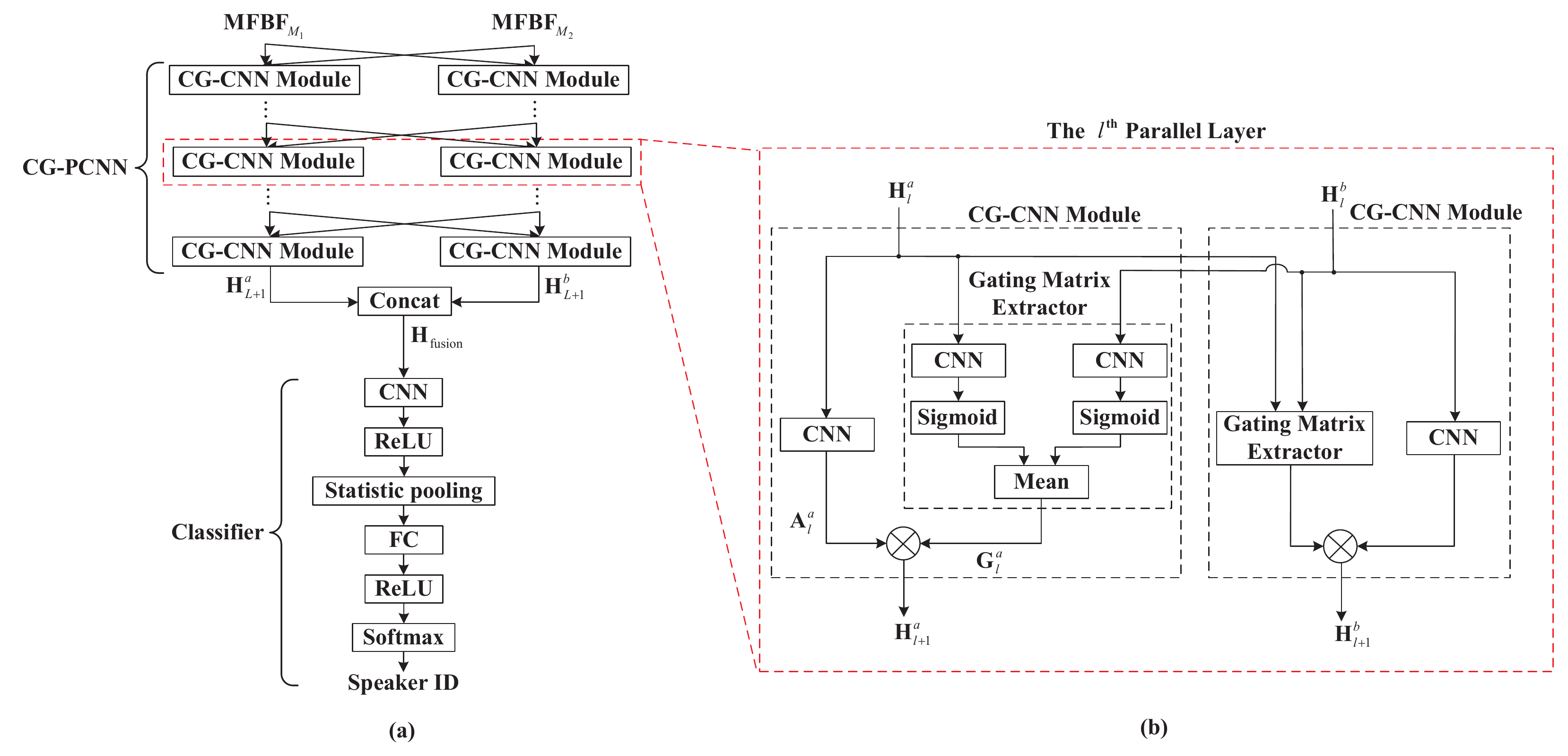}
		\caption{The proposed speaker recognition system. (a) represents the structure of the speaker recognition system based on the CG-PCNN, (b) represents the structure of the $l^{th}$ parallel layer in the CG-PCNN.}
		\label{f2}
	\end{figure*}
	The $M$-dimensional power spectrum can be extracted from the speech frame by the Mel filter bank. Then, the $M$-dimensional MFBF is obtained by computing the log of the $M$-dimensional power spectrum  of the speech frame and denoted by MFBF$_M$.
	
	Generally, $M$ can be selected as 13, 26 and 40, respectively, and the frequency domain responses of the Mel filter banks are shown in Fig. \ref{f3} (a), (b) and (c). As $M$ is increased, the bandwidth of each triangular filter in the Mel filter bank is decreased and the resolution of the frequency domain of the Mel filter bank is higher.
	As shown in Fig. \ref{f4} (a), (b) and (c), the MFBF$_{13}$, MFBF$_{26}$ and MFBF$_{40}$  are extracted from a speaker's speech, respectively. The areas between the two red dashed lines in the three feature maps represent the same frequency range of the speech signals, i.e., approximately the 320 Hz $ \sim $  1500 Hz frequency range. In this frequency range, the dimensions of the MFBF$_{13}$, MFBF$_{26}$ and MFBF$_{40}$ are 3, 7 and 12, respectively. In the same frequency range, when $M$ is increased, the frequency resolution of the MFBF$_M$ is higher, and more details in the frequency domain is contained in the MFBF$_M$. Thus, there are some complementarities  between the MFBF$_{M_1}$ and MFBF$_{M_2}$ of a speech frame in the same frequency range, where the ${M_1}$ and ${M_2}$ are different.
	
	\section{Proposed Feature Fusion Method}
	The structure of the proposed speech feature fusion method is shown in Fig. \ref{f2} (a).
	The cross gate parallel convolutional neural network (CG-PCNN) is used to select the efficient features from the MFBF$_{M_1}$ and MFBF$_{M_2}$ of a speaker's speech, respectively.
	Then, the outputs of the CG-PCNN are concatenated as the fusion feature of a speaker's speech, and the fusion feature is fed into the classifier to predict the identity of the speaker.
	
	\subsection{Fusion Feature Extraction}
	The structure of the CG-PCNN is demonstrated in Fig. \ref{f2} (a). The CG-PCNN consists of $L$ parallel layers, and each parallel layer contains two cross gate CNN modules (CG-CNN modules) . Let ${\bf{H}}_1^a \in {\mathbb{R}^{{M_1} \times T}}$ and ${\bf{H}}_1^b \in {\mathbb{R}^{{M_2} \times T}}$ denote the MFBF$_{M_1}$ and MFBF$_{M_2}$ of a speaker's speech, and are employed as the inputs of the CG-PCNN, respectively. $M_1$ and $M_2$ are the dimensions of the MFBF$_{M_1}$ and MFBF$_{M_2}$,  respectively. $T$ is the number of frames of the speaker's speech.
	
	As shown in Fig. \ref{f2} (b),
	${\bf{H}}_l^a \in {\mathbb{R}^{D_l^a \times {T_l}}}$, ${\bf{H}}_l^b \in {\mathbb{R}^{D_l^b \times {T_l}}}$ are employed as the inputs of the $l^{th} {, \ _{l=1,2,...,L}}$ parallel layer,  respectively. $D_l^a$ and $D_l^b$ denote the hights of the ${\bf{H}}_l^a$ and ${\bf{H}}_l^b$, respectively. ${T_l}$ represents the width of the ${\bf{H}}_l^a$ and ${\bf{H}}_l^b$.
	
	The left CG-CNN module of the ${l^{th}}$ parallel layer consists of a CNN layer and a gating matrix extractor, which also can be seen as a CNN layer with a cross gate mechanism.
	The CNN layer is utilized to grab out the essential information hidden in ${\bf{H}}_l^a$:
	\begin{equation}
		{\bf{A}}_l^{a} = {{Con}}v\left( {{\bf{H}}_l^a,\bm\uptheta _l^a} \right)
	\end{equation}
	where ${{\bf{A}}_l^{a}}  \in {\mathbb{R}^{D_{l+1} \times T_{l+1}}}$ is the output of the CNN layer. $D_{l+1}$ and  $T_{l+1}$ are the height and width of ${{\bf{A}}_l^{a}}$, respectively. $Conv\left(  \cdot  \right)$ denotes the convolutional operation of the CNN layer and ${\bm\uptheta _l^a}$ represents the parameters of the CNN layer.
	
	The gating matrix extractor in the left CG-CNN module of the $l^{th}$ parallel layer is applied to obtain the complementaries within the MFBF$_{M_1}$ and MFBF$_{M_2}$ of a speaker's speech, and the complementaries are represented by a gating weight matrix. ${\bf{H}}_l^a$ and ${\bf{H}}_l^b$ are utilized as the inputs of two CNN layers, respectively, where the outputs of the CNN layers are activated by the sigmoid activation function. The gating weight matrix ${{\bf{G}}_l^a}  \in {\mathbb{R}^{D_{l+1} \times T_{l+1}}}$ is the mean of the  outputs of the two CNN layers :	
	\begin{equation}
		{\bf{G}}_l^a = {{( {\delta ( {{\mathop{\rm Conv}\nolimits} ( {{\bf{H}}_l^a,{\bm{\uptheta }}_l^{aa}} )} ) + \delta ( {{\mathop{\rm Conv}\nolimits} ( {{\bf{H}}_l^b,{\bm{\uptheta }}_l^{ba}} )} )} )} \mathord{\left/
				{\vphantom {{( {\delta ( {{\mathop{\rm Conv}\nolimits} ( {{\bf{H}}_l^a,{\bm{\uptheta }}_l^{aa}} )} ) + \delta ( {{\mathop{\rm Conv}\nolimits} ( {{\bf{H}}_l^b,{\bm{\uptheta }}_l^{ba}} )} )} )} 2}} \right.
				\kern-\nulldelimiterspace} 2}
	\end{equation}
	where $\delta  \left(  \cdot  \right)$ represents the sigmoid activation function. ${\bm\uptheta _l^{aa}}$ and ${\bm\uptheta _l^{ba}}$ are the parameters of  the CNN layers, respectively.
	
	Then, the feature ${\bf{A}}_l^{a}$ is enhanced by the cross gate mechanism with the complementaries within the MFBF$_{M_1}$ and MFBF$_{M_2}$ of a speaker's speech:
		\begin{equation}
			{\bf{H}}_{l + 1}^a = {\bf{A}}_l^{a}  \otimes  {{\bf{G}}_l^a}
		\end{equation}
	where ${{\bf{H}}_{l+1}^a}  \in {\mathbb{R}^{D_{l+1} \times T_{l+1}}}$ is the output of the left CG-CNN module in the $l^{th}$ parallel layer, $\otimes$ means the element-wise product.
	
	The structures of the right and left CG-CNN modules in the $l^{th}$ parallel layer of the CG-PCNN are symmetric, and the output feature ${{\bf{H}}_{l+1}^b}  \in {\mathbb{R}^{D_{l+1} \times T_{l+1}}}$ of the right CG-CNN module can be written as:
	\begin{table*}[!t]
		\renewcommand{\arraystretch}{1.3}
		\caption{The architecture of the speaker recognition system based on the CG-PCNN: $M_1$ and $M_2$ are dimensions of the two MFBFs, respectively. The dilation method \cite{J36} is conducted in each CNN layer of each CG-CNN module.}
		\label{t1}
		\centering
		\begin{tabular}{ccccccc}
			\hline
			Layer&Network &Kernel Size&	Kernel Size & Stride & Dilation &Output Size\\
			\hline
			1&CG-CNN module	&$ {M_1} \times 5 \times 256$&$ {M_2} \times 5 \times 256$&1&1&	$256 \times 296\times 2$\\
			2&CG-CNN module&	$256 \times 5 \times 256$&$256 \times 5 \times 256$&1&2&	$256 \times 288\times 2$\\
			3&CG-CNN module&	$256 \times 7 \times 256$&$256 \times 7 \times 256$&1&3&	$256 \times 270\times 2$\\
			4&CG-CNN module&	$256 \times 1 \times 256$&$256 \times 1 \times 256$&1&1&	$256 \times 270\times 2$\\
			5&Concatenation &\multicolumn{2}{c}{-}&-&-&$512 \times 270$\\
			6&CNN&	\multicolumn{2}{c}{$512 \times 1 \times 1500$}&1&1&	$1500 \times 270$\\
			7&Statistics pooling&\multicolumn{2}{c}{-}&-&-&	$3000 \times 1$\\
			8&FC	&\multicolumn{2}{c}{-}&-&-&	$512$\\
			9&Softmax&\multicolumn{2}{c}{-}&-&-&	$Number of the Speakers$\\
			\hline
		\end{tabular}
	\end{table*}
	\begin{equation}
		\begin{array}{l}
			{\bf{H}}_{l + 1}^b{\rm{ = }}Conv( {{\bf{H}}_l^b,\bm\uptheta _l^b} )\\
			\quad\quad \otimes {{( {\delta ( {{{Con}}v( {{\bf{H}}_l^b,\bm\uptheta _l^{bb}} )} ) + \delta ( {{{Con}}v( {{\bf{H}}_l^a,\bm\uptheta _l^{ab}} )} )} )} \mathord{\left/
					{\vphantom {{( {\delta ( {{{Con}}v( {{\bf{H}}_l^b,\bm\uptheta _l^{bb}} )} ) + \delta ( {{{Con}}v( {{\bf{H}}_l^a,\bm\uptheta _l^{ab}} )} )} )} 2}} \right.
					\kern-\nulldelimiterspace} 2}
		\end{array}
	\end{equation}
	where ${\bm\uptheta _l^{b}}$, ${\bm\uptheta _l^{bb}}$ and ${\bm\uptheta _l^{ab}}$ are the parameters of the CNN layers in the ${l^{th}}$ parallel layer of the CG-PCNN, respectively.
	
	Then, the fusion feature can be obtained:
	\begin{equation}{{\bf{H}}_{\rm{fusion}}} = \left[ {\begin{array}{*{20}{c}}
				{{\bf{H}}_{{L+1}}^{a}}\\
				{{\bf{H}}_{{L+1}}^{b}}
		\end{array}} \right]
	\end{equation}
	where ${\bf{H}}_{L+1}^a$ and ${\bf{H}}_{L+1}^b$ are the outputs of the $L^{th}$ parallel layer of the CG-PCNN, respectively.
	
	\subsection{Classifier}
	In the classifier, the efficient deep feature is selected from the fusion feature ${\bf{H}}_{\rm{fusion}}$ by a CNN layer, where the output of the CNN layer is activated by the ReLU activation function:
		\begin{equation}
		{\bf{H}} = \tau ({{Con}}v\left( {{\bf{H}}_{\rm{fusion}},\bm\uptheta} \right))
	\end{equation}
	where ${\bf{H}}{ \in \mathbb{R}^{D' \times T'}}$. $D'$ and $T'$ are the height and width of ${\bf{H}}$, respectively. $\tau$ and $\bm\uptheta$ denote the ReLU activation function and the parameters of the CNN layer, respectively.
	
	Then, ${\bf{H}}$ is aggregated into a statistic feature by a statistics pooling layer \cite{J23}:
	\begin{equation}
		{\mu _i} = \frac{1}{{T'}}\sum\limits_{t = 0}^{T' - 1} {{H_{i,t}}} , \ i = 0,1, \ldots ,D' - 1
	\end{equation}
	\begin{equation}
		{\sigma _i} = \sqrt {\frac{1}{{T'}}\sum\limits_{t = 0}^{T' - 1} {{{\left( {{H_{i,t}} - {\mu _i}} \right)}^2}} } , \ i = 0,1, \ldots ,D' - 1
	\end{equation}
	where ${H_{i,t}}$ represents the element in the $i^{th}$ row and the $t^{th}$ column of $\bf{H}$.
	The mean vector ${\bm{\upmu }} = \left[ {{\mu _0}, \ldots ,{\mu _{D' - 1}}} \right]{ \in \mathbb{R}^{D' \times 1}}$ and the standard deviation vector ${\bm{\upsigma }} = \left[ {{\sigma _0}, \ldots ,{\sigma _{D' - 1}}} \right]{ \in \mathbb{R}^{D' \times 1}}$ are concatenated as the statistic feature ${\bf{s}} = {\left[ {{{\bm{\upmu }}^{\rm{T}}},{{\bm{\upsigma }}^{\rm{T}}}} \right]^{\rm{T}}} \in {\mathbb{R}^{2D' \times 1}}$.

	Finally, the statistics feature ${\bf{s}}$ is fed into a fully connected layer for dimension reduction, where the output of the fully connected layer is activated by the ReLU activation function. And the output of the fully connected layer is classified by a softmax layer for speaker recognition.	
	
	\section{Experiments}
	\subsection{Datasets and Experimental Setups}
	The AISHELL-1 \cite{J50}, Librispeech \cite{J53} and Voxceleb2 \cite{J26} datasets are used to evaluate the performance of the proposed speaker recognition system. In this paper, 100 speakers are randomly selected from the AISHELL-1 dataset and 251 speakers are randomly selected from the Librispeech dataset. Each of the speaker has 40 sentences. Among them, 20 sentences of each speaker are selected for training, and another 20 sentences of each speaker are used for testing. In addition, 100 speakers are randomly selected from the Voxceleb2 dataset, and each of the speaker has 120 sentences. 100 sentences of each speaker are selected for training, and another 20 sentences of each speaker are utilized for testing.

	For all the sentences of each speaker, the length of each sentence is 3 seconds and is divided into 25 ms length frames by using a sliding hamming window with 10 ms of overlap. Then, the MFBF$_{13}$, MFBF$_{26}$ and MFBF$_{40}$ are extracted from each sentence, respectively. In addition, the MFBF$_{13}$, MFBF$_{26}$ and MFBF$_{40}$ of each sentence are normalized with the cepstral mean normalization (CMN) \cite{J31}, respectively.

		\begin{figure*}[!t]
		\centering
		\includegraphics[width=5.5in]{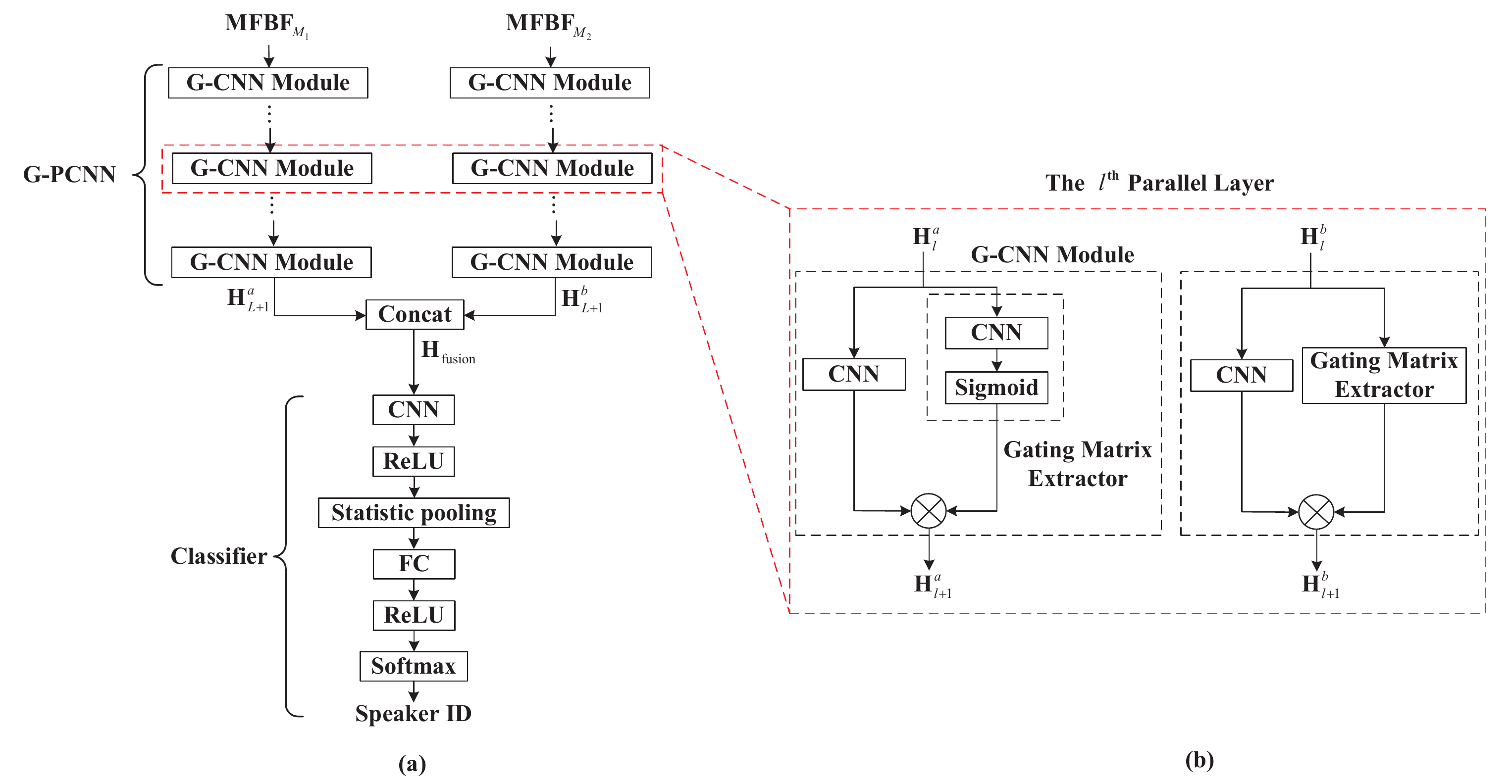}
		\caption{The speaker recognition system based on the G-PCNN. (a) represents the structure of  speaker recognition system based on the G-PCNN, (b) represents the structure of the $l^{th}$ parallel layer in the G-PCNN, which consists of two G-CNN modules \cite{J39}, \cite{J57}.}
		\label{f5}
		\end{figure*}
	
	The architecture of the CG-PCNN is shown in TABLE \ref{t1} and  the network is trained with the Adam optimizer. The epoch is selected as 200 and the learning rate is decaying from ${10^{ - 3}}$ to ${10^{ - 4}}$ by experiments.

	\begin{table*}[!t]
		\renewcommand{\arraystretch}{1.3}
		\caption{The experimental results of the speaker recognition systems on the AISHELL-1, Librispeech and Voxceleb2 datasets, the MFBF$_{M_1}$$\&$MFBF$_{M_2}$ respesent the MFBF$_{M_1}$ and MFBF$_{M_2}$ of a speaker's speech
		}
		\label{t2}
		\centering
		\begin{tabular}{cccccccc}
			\hline
			\multirow{2}{*}{Network}&\multirow{2}{*}{Feature} & \multicolumn{2}{c}{AISHELL-1}& \multicolumn{2}{c}{Librispeech}& \multicolumn{2}{c}{Voxceleb2}\\
			\cline{3-8}
			&& Mean (\%)& STD& Mean (\%)& STD& Mean (\%)& STD\\
			\hline
			\multirow{3}{*}{CG-PCNN}
			&MFBF$_{13}$$\&$MFBF$_{26}$& 98.80& 0.19& 96.86& 0.12& 86.35& 0.83\\
			&MFBF$_{13}$$\&$MFBF$_{40}$& 99.03& 0.20& 97.02& 0.13& \bf{86.88}& 0.97\\
			&MFBF$_{26}$$\&$MFBF$_{40}$& \bf{99.23}& 0.15& \bf{97.05}& 0.12& 86.44& 0.53\\
			\hline
			\multirow{3}{*}{PCNN}
			&MFBF$_{13}$$\&$MFBF$_{26}$& 98.41& 0.23& 96.28& 0.22& \bf{84.90}& 0.93\\
			&MFBF$_{13}$$\&$MFBF$_{40}$& 98.61& 0.17& \bf{96.65}& 0.14& 84.80& 0.95\\
			&MFBF$_{26}$$\&$MFBF$_{40}$& \bf{98.72}& 0.27& 96.58& 0.06& 84.54& 0.96\\
			\hline
			\multirow{3}{*}{G-PCNN}
			&MFBF$_{13}$$\&$MFBF$_{26}$& 98.71& 0.15& 96.41& 0.14& 85.51& 0.77\\
			&MFBF$_{13}$$\&$MFBF$_{40}$& 98.93& 0.25& 96.67& 0.16& \bf{85.66}& 0.58\\
			&MFBF$_{26}$$\&$MFBF$_{40}$& \bf{99.07}& 0.13& \bf{96.76}& 0.17& 85.41& 0.77\\
			\hline
			\multirow{3}{*}{SFAN}&MFBF$_{13}$& 94.65& 0.36& 94.42& 0.20& 80.87& 0.60\\
			&MFBF$_{26}$& 97.83& 0.36& 95.77& 0.14& \bf{83.91}& 0.68\\
			&MFBF$_{40}$& \bf{98.00}& 0.30& \bf{96.12}& 0.23& 83.40& 0.51\\
			\hline
		\end{tabular}
	\end{table*}

	\begin{table*}[!t]
		\renewcommand{\arraystretch}{1.3}
		\caption{The experimental results of the speaker recognition systems under the white noise background conditions on the AISHELL-1 dataset
		}
		\label{t3}
		\centering
		\setlength{\tabcolsep}{3mm}{
			\begin{tabular}{cccccc}
				
				\hline
				\multirow{2}{*}{Network}& \multirow{2}{*}{Feature}& \multicolumn{2}{c}{25dB}&\multicolumn{2}{c}{30dB}\\
				\cline{3-6}
				&& Mean (\%)& STD& Mean (\%)& STD\\
				\hline
				\multirow{3}{*}{CG-PCNN}
				&MFBF$_{13}$$\&$MFBF$_{26}$& 65.36& 3.06& 88.05& 0.67\\
				&MFBF$_{13}$$\&$MFBF$_{40}$& \bf{68.38}& 2.67& \bf{89.26}& 1.08\\
				&MFBF$_{26}$$\&$MFBF$_{40}$& 68.20& 1.78& 88.63& 0.90\\
				\hline
				\multirow{3}{*}{PCNN}
				&MFBF$_{13}$$\&$MFBF$_{26}$& 63.99& 3.20& 86.51& 1.30\\
				&MFBF$_{13}$$\&$MFBF$_{40}$& 67.57& 2.20& \bf{87.47}& 0.83\\
				&MFBF$_{26}$$\&$MFBF$_{40}$& \bf{67.93}& 2.76& 87.44& 1.12\\
				\hline
				\multirow{3}{*}{G-PCNN}
				&MFBF$_{13}$$\&$MFBF$_{26}$& 62.13& 1.80& 86.10& 0.86\\
				&MFBF$_{13}$$\&$MFBF$_{40}$& 64.71& 2.72& 86.94& 1.60\\
				&MFBF$_{26}$$\&$MFBF$_{40}$& \bf{65.69}& 1.54& \bf{87.24}& 1.14\\
				\hline
				\multirow{3}{*}{SFAN}&MFBF$_{13}$& 45.15& 1.44& 72.39& 1.32\\
				&MFBF$_{26}$& 57.72& 3.22& 82.15& 2.00\\
				&MFBF$_{40}$& \bf{60.36}& 2.65& \bf{82.31}& 1.48\\
				\hline
		\end{tabular}}
	\end{table*}
	
	\subsection{Performance of the Proposed Speaker Recognition System}
	To evaluate the performance of the proposed speaker recognition system based on the CG-PCNN, two of the MFBF$_{13}$, MFBF$_{26}$ and MFBF$_{40}$ of a speaker's speech are selected as the inputs of the proposed speaker recognition system, respectively, e.g., the MFBF$_{13}$$\&$MFBF$_{26}$, MFBF$_{13}$$\&$MFBF$_{40}$ and MFBF$_{26}$$\&$MFBF$_{40}$ of the speaker's speech. The experiment is conducted for 10 times, the results of the mean and standard deviation (STD) of the speaker recognition rates (SRRs)  are shown in Table \ref{t2}.
	For the AISHELL-1 dataset, the SRR of the speaker recognition system with the MFBF$_{26}$$\&$MFBF$_{40}$ is 99.23\%, and it is higher than the SRRs of the speaker recognition system with the MFBF$_{13}$$\&$MFBF$_{26}$ and  MFBF$_{13}$$\&$MFBF$_{40}$, respectively. For the Librispeech dataset, the SRR of the speaker recognition system with the  MFBF$_{26}$$\&$MFBF$_{40}$ is 97.05\%, it is also higher than the SRRs of the speaker recognition system with the MFBF$_{13}$$\&$MFBF$_{26}$ and  MFBF$_{13}$$\&$MFBF$_{40}$, respectively. For the Voxceleb2 dataset, the SRR of the speaker recognition system with the MFBF$_{13}$$\&$MFBF$_{40}$ is 86.88\%, and it is higher than the SRRs of the speaker recognition system with the MFBF$_{13}$$\&$MFBF$_{26}$ and MFBF$_{26}$$\&$MFBF$_{40}$, respectively. The experimental results show that the speaker recognition system with the MFBF$_{26}$$\&$MFBF$_{40}$ achieves the best performance for the AISHELL-1 and Librispeech datasets, and the speaker recognition system with the MFBF$_{26}$$\&$MFBF$_{40}$ only achieves the best performance for the Voxceleb2 dataset. These illustrate that the proposed speaker recognition system with the MFBF$_{26}$$\&$MFBF$_{40}$ is more suitable than the proposed speaker recognition system with the  other inputs for speaker recognition.
	
	In the CG-PCNN, each of the CG-CNN module is a CNN layer with a cross gate mechanism. To evaluate the effectiveness of the cross gate mechanism, we construct the parallel convolutional neural network (PCNN), which utilizes the same structure as the CG-PCNN, and each of the CG-CNN module is replaced by one CNN layer with the ReLU activation function. The results of the mean and STD of the SRRs for the experiment on the PCNN are shown in Table \ref{t2}. For the  Voxceleb2 dataset,
	when the inputs of each network are the MFBF$_{13}$$\&$MFBF$_{26}$, MFBF$_{13}$$\&$MFBF$_{40}$ and MFBF$_{26}$$\&$MFBF$_{40}$, the SRRs of the speaker recognition system based on the CG-PCNN are superior to the SRRs of the speaker recognition system based on the PCNN by 1.45\%, 2.08\% and 1.90\%, respectively.
    The experimental results show that the speaker recognition system based on the CG-PCNN is outperform the speaker recognition system based on the PCNN, when the inputs of the PCNN and CG-PCNN are the same.
	For the AISHELL-1 and Librispeech datasets, the same phenomenon can be also found in Table \ref{t2}. These illustrate that the cross gate mechanism is effective for speaker recognition and can be used to improve the performance of the speaker recognition system.
	
	Similar to the CG-CNN module, the gate convolutional neural network module (G-CNN module) \cite{J39}, \cite{J57} is a CNN layer with a gate mechanism, and it could be formulated as:
		\begin{equation}
		{{\bf{H}}_{{\rm{out}}}} = Conv\left( {{{\bf{H}}_{{\rm{in}}}},\bm\uptheta } \right) \otimes \delta \left( {Conv\left( {{{\bf{H}}_{{\rm{in}}}},{\bm\uptheta _{\rm{g}}}} \right)} \right)
	\end{equation}
	where ${{\bf{H}}_{\rm{out}}}$ and ${{\bf{H}}_{\rm{in}}}$ are the output and input of the G-CNN module, respectively. $Conv\left(  \cdot  \right)$ denotes the convolutional operation of the CNN layer. ${\bm\uptheta}$ and ${\bm\uptheta _{\rm{g}}}$ represents the parameters of the CNN layers.
	As shown in Fig. \ref{f5}, we construct the gate parallel convolutional network (G-PCNN), which employs the same structure as the CG-PCNN, and all the CG-CNN modules are replaced by the gate CNN modules.
	The results of the mean and STD of the SRRs for the experiment on the G-PCNN are shown in Table \ref{t2}.
	Comparing the SRR of the speaker recognition system based on the CG-PCNN and the SRR of the speaker recognition system based on the G-PCNN, we can see that the speaker recognition system based on the CG-PCNN outperforms the speaker recognition system based on the G-PCNN, where the inputs of the CG-PCNN and G-PCNN are the same. For instance,
	when the inputs of each network are the MFBF$_{13}$$\&$MFBF$_{40}$,
	the SRR of the speaker recognition system based on the CG-PCNN is superior to the SRR of the speaker recognition system based on the G-PCNN by 1.22\% for the Voxceleb2 dataset.
	It illustrates that the complementarities between the features of different frequency resolutions can be employed by the cross gate mechanism to improve the performance of a speaker recognition system. The proposed feature fusion method based on the CG-PCNN is effective for speaker recognition.
	
	Except for using the fusion feature of a speaker's speech, only one kind of the MFBF$_M$ of a speaker's speech can also be used for speaker recognition. To employ the MFBF$_M$ for speaker recognition, a single feature analysis network (SFAN) is utilized, which removes one CNN branch and  the concatenation operation of the P-CNN. The MFBF$_{13}$, MFBF$_{26}$ and MFBF$_{40}$ are employed as the input of the SFAN, respectively. The results of the mean and STD of the SRRs for the experiment on the SFAN are shown in Table \ref{t2}.
	We compare the performance of the speaker recognition system based on the SFAN with the performance of the speaker recognition system based on the CG-PCNN.
	When the MFBF$_{13}$, MFBF$_{26}$ and MFBF$_{40}$ are employed as the input of the SFAN, respectively, the SRRs of the speaker recognition system based on the SFAN are 94.65\%, 97.83\% and 98.00\% for the AISHELL-1 dataset. For the Librispeech dataset, the SRRs of the speaker recognition system based on the SFAN are 94.42\%, 95.77\%, 96.12\%, respectively. For the Voxceleb2 dataset, the SRRs of the speaker recognition system based on the SFAN are 80.87\%, 83.91\%, 83.40\%, respectively. For the AISHELL-1, Librispeech and Voxceleb2 datasets, the highest SRRs of the speaker recognition system based on the SFAN are 98.00\%, 96.12\% and 83.91\%, respectively. But they are inferior to the lowest SRRs of the speaker recognition system based on the CG-PCNN by 0.80\%, 0.72\% and 2.44\%, respectively. It illustrates that the feature fusion method based on the CG-PCNN is effective for speaker recognition, and the complementarities between the features of different frequency resolutions can be utilized to improve the performance of the speaker recognition system.
	
	\subsection{Effectiveness of the Proposed Speaker Recognition System under the White Noise Background Conditions}
	To evaluate the performance of the proposed speaker recognition system under the white noise background conditions, the extra white noises are artificially added into the test speeches in the AISHELL-1 dataset. The signal to noise ratios (SNRs) of the test speeches are 25dB and 30dB, respectively. The speaker recognition systems based on the CG-CNN, PCNN and G-PCNN are tested on these test speeches, respectively. The mean and STD of the SRRs for each experiment are shown in Table \ref{t3}.
	Where the inputs of the CG-PCNN are the MFBF$_{13}$$\&$MFBF$_{26}$, the SRRs of the proposed speaker recognition system are 65.36\% and 88.05\%, respectively, when the SNRs of the test speeches are 25dB and 30dB. And the SRRs of the proposed speaker recognition system are superior to the SRRs of the speaker recognition system based on the PCNN by 1.37\% and 1.54\%, respectively. The SRRs of the proposed speaker recognition system are superior to the SRRs of the speaker recognition system based on the G-PCNN by 3.23\% and 1.95\%, respectively. When the inputs of each network are the MFBF$_{13}$$\&$MFBF$_{26}$, the SRR of the proposed speaker recognition system is higher than the SRRs of the speaker recognition systems based on the PCNN and G-PCNN, respectively. And the same phenomenon can be found where the inputs of each network are the MFBF$_{13}$$\&$MFBF$_{40}$ or the MFBF$_{26}$$\&$MFBF$_{40}$.
	These results illustrate that the proposed speaker recognition system is more robust than the speaker recognition systems based on the PCNN and G-PCNN, respectively, under the white noise background conditions.

	To compare the performance of the speaker recognition system based on the CG-PCNN with the performance of the speaker recognition based on the SFAN under the white noise background conditions, the extra white noises are added into the test speeches in the AISHELL-1 dataset, and the SNRs of the test speeches are 25dB and 30dB, respectively. The speaker recognition systems based on the CG-PCNN and SFAN are tested on these test speeches, respectively.
	The mean and STD of the SRRs for each experiment are shown in Table \ref{t3}.
	When the SNRs of each test speech are 25dB and 30dB, the highest SRRs of the speaker recognition based on the SFAN are 60.36\% and 82.31\%, respectively, but they are inferior to the lowest SRRs of the speaker recognition system based on the CG-PCNN by 5.00\% and 5.74\%, respectively. These points prove that the fusion feature extracted by the CG-PCNN is more robust than the single feature based on the SFAN, and the proposed feature fusion method is effective under white the noise background conditions.
	
	\subsection{Comparation with Baseline Systems}
	To evaluate the performance of the proposed speaker recognition system, five baseline systems are employed.
	\subsubsection{i-vector \cite{J21}}
	Based on the 13-dimensional mean and variance normalized GFCC of speaker's speech, the total variability algorithm is used to extract the i-vector feature, then the i-vector feature is utilized as the input of the speaker's PLDA model for speaker recognition.
	
	\subsubsection{x-vector \cite{J23}}
	The x-vector feature extraction system employs the MFBF$_{24}$ of a speaker's speech as the input of the DNN  for speaker recognition, where the DNN contains five TDNN layers, one statistics pooling layer, two fully connected layers and one softmax layer.
	
	\subsubsection{Spectrogram+CNN \cite{J16}}
	The 512-dimensional spectrogram feature is extracted from a speaker's speech, then the cepstral mean and variance normalization (CMVN) is applied on the spectrogram feature of the speaker's speech. Finally, the spectogram feature is employed as the input of a CNN for speaker recognition.
	
	\subsubsection{MFCC+CNN \cite{J43}}
	The MFCC+CNN employs the concatenation of the 40-dimensional MFCC and 40-dimensional delta-MFCC of a speaker's speech as the input of a CNN for speaker recognition, where the CNN contains five convolutional layers and two fully connected layers.
	
	\subsubsection{MFCC+CENS+CNN \cite{J43}}
	The MFCC+CENS+CNN employs the concatenation of the 40-dimensional MFCC, 40-dimensional delta-MFCC and 12-dimensional CENS of a speaker's speech as the fusion feature. Then the fusion feature of the speaker's speech is employed as the input of the CNN for speaker recognition, where the CNN contains five convolutional layers and two fully connected layers.
	
	The CG-PCNN with the MFBF$_{26}$$\&$MFBF$_{40}$ of a speaker's speech is employed as the proposed speaker recognition system.
	The experimental results are shown in Table \ref{t4}.
	For the  AISHELL-1 dataset, the SRR of the proposed speaker recognition system is superior to the SRRs of these baseline systems by 4.32\% on the i-vector, 0.78\% on the x-vector, 26.97\% on the Spectrogram+CNN, 12.45\% on the MFCC+CNN and 12.13\% on the MFCC+CENS+CNN, respectively.
	For the Librispeech dataset, the SRR of the
		\begin{table}[htbp]
		\renewcommand{\arraystretch}{1.3}
		\caption{The experimental results on the AISHELL-1, Librispeech, Voxceleb2 datasets with the baseline systems and the proposed speaker recognition system}
		\label{t4}
		\centering
		\setlength{\tabcolsep}{0.6mm}{
			\begin{tabular}{ccccccc}
				\hline
				\multirow{2}{*}{Method}& \multicolumn{2}{c}{AISHELL-1}& \multicolumn{2}{c}{Librispeech}& \multicolumn{2}{c}{Voxceleb2}\\
				\cline{2-7}
				& Mean (\%)& STD& Mean (\%)& STD& Mean (\%)& STD\\
				\hline
				i-vector \cite{J21}&	96.14&	0.08&96.64&0.10& 66.21& 0.50\\
				
				x-vector \cite{J23}&	98.45&	0.24&96.36&0.22& 85.26& 0.66\\
				
				Spectrogram+CNN \cite{J16}	&	72.26&	1.17&64.40&1.62& 66.35& 0.78\\
				
				MFCC+CNN \cite{J43}  &	86.78&	0.57&91.50&0.40& 60.59& 1.07\\
				
				MFCC+CENS+CNN \cite{J43}  &	87.10&	0.91&91.54&0.72& 61.07& 1.07\\
				
				CG-PCNN&	\bf{99.23}&	0.15&\bf{97.05}&0.12& \bf{86.44}& 0.58\\
				\hline
		\end{tabular}}
	\end{table}
	\begin{table}[htbp]
		\renewcommand{\arraystretch}{1.3}
		\caption{The experimental results of the baseline systems and the proposed speaker recognition system under the white noise background conditions on the AISHELL-1 dataset}
		\label{t5}
		\centering
		\setlength{\tabcolsep}{1.5mm}{
			\begin{tabular}{ccccc}
				\hline
				\multirow{2}{*}{Method}& \multicolumn{2}{c}{25dB}&\multicolumn{2}{c}{30dB}\\
				\cline{2-5}
				& Mean (\%)& STD& Mean (\%)& STD\\
				\hline
				i-vector \cite{J21}& 63.88& 0.18& 84.55& 0.23\\
				x-vector \cite{J23}& 56.50& 2.89& 82.20& 1.79\\
				Spectrogram+CNN \cite{J16} & 17.76& 1.78 & 32.83& 1.94\\
				MFCC+CNN \cite{J43}  & 40.83& 3.65& 58.44& 4.02\\
				MFCC+CENS+CNN \cite{J43}  & 39.14& 5.47& 57.52& 4.81\\
				CG-PCNN& \bf{68.20}& 1.78& \bf{88.63}& 0.90\\
				\hline
		\end{tabular}}
	\end{table}
	speaker recognition systems based on the CG-PCNN is superior to the SRRs of these baseline systems by 0.41\% on the i-vector, 0.69\% on the x-vector, 32.65\% on the Spectrogram+CNN, 5.55\% on the MFCC+CNN and 5.51\% on the MFCC+CENS+CNN, respectively.
	For the Voxceleb2 dataset, the SRR of the speaker recognition systems based on the CG-PCNN is   superior to the SRRs of these baseline systems by 20.23\% on the i-vector, 1.18\% on the x-vector, 20.09\% on the Spectrogram+CNN, 25.85\% on the MFCC+CNN and 25.37\% on the MFCC+CENS+CNN, respectively.
	The experimental results show that  the proposed speaker recognition system achieves best performance among all the speaker recognition systems shown in Table \ref{t4}.
	
	As shown in Table \ref{t5}, the proposed speaker recognition system achieves the highest SRR among all the speaker recognition systems, when the SNRs of the test speeches are 25dB and 30dB, respectively. These illustrate the proposed speaker recognition system is more robust than the x-vector, the Spectrogram+CNN, the MFCC+CNN and the MFCC+CENS+CNN under the white noise background conditions. The proposed speaker recognition systems also achieve the comparable results with the i-vector under the white noise background conditions.
	
	\section{Conclusion}
	\label{sec:print}
	In this paper, we propose a speech feature fusion method based on the CG-PCNN to fuse the MFBFs of different frequency resolutions for speaker recognition.
	By using the CG-PCNN, the complementaries within the MFBFs of different frequency resolutions of each speech frame are obtained for speaker recognition, and the performance of the speaker recognition system is improved. In addition, the five baseline systems (the i-vector, x-vector, Spectrogram+CNN, MFCC+CNN and MFCC+CENS+CNN) are utilized for experimental comparison. The speaker recognition system with the proposed feature fusion method is robust and outperforms all the baseline systems.
	
	\bibliographystyle{IEEEtran}
	
	\bibliography{IEEEabrv,mybibfile}
	\begin{IEEEbiographynophoto}{Jiacheng Zhang}
		is currently pursuing the master’s degree at Nanchang University, Nanchang, China. His current research focuses on speaker recognition.
	\end{IEEEbiographynophoto}
	\begin{IEEEbiographynophoto}{Wenyi Yan}
		is currently pursuing the master’s degree at Nanchang University, Nanchang, China. His current research focuses on speaker recognition.
	\end{IEEEbiographynophoto}
	\begin{IEEEbiographynophoto}{Ye Zhang}
		received the Ph.D. degree in information and communication engineering from Shanghai University, Shanghai, China, in 2009. He is currently a Professor at Nanchang University, Nanchang, China. His current research interests include blind signal processing, speech and image signal processing, pattern recognition and machine learning.
	\end{IEEEbiographynophoto}
	\vfill

\end{document}